\documentclass[prb,amsmath,amssymb]{revtex4}

\begin{document}


\title{Exact, explicit and entire solutions to a nontrivial finite-difference equation and their applications}


\author{M. Aunola}
\affiliation{Department of Physics, University of Jyv\"askyl\"a,
P.O. Box 35 (YFL), FIN-40014 University of Jyv\"askyl\"a, Finland}


\date{\today}

\begin{abstract}
The explicit solution to a certain finite-difference equation is given and the required steps for derivation of these results are outlined.  Everything is included as MATHEMATICA formulae, so the notebook itself
can be used for checking and improving the present results. Some important references for justifying some steps and crosschecking certain results have been included. Full references and derivations will be made available shortly. It should be noted that several applications for the solutions have been included at the end of the document. These include at least diagonalisation of certain infinite matrices, definition of isospectral operators with simple eigenvalues and alternative representation of a solution to a problem related to black holes. Additionally, the corresponding discretised associated Laguerre polynomials  are defined explicitly.
\end{abstract}


\maketitle


\section{Introduction\label{sec:intro}}

Explicit, exact and entire solutions to a finite-difference equation
\begin{equation}
-u(z-\delta)/2-u(z+\delta)/2-\delta^2 u(z)/z=\lambda^{(\delta)} u(z)
\label{eq:finite}
\end{equation}
under the condition $u(0)=0$ have been found. Alternatively, the functions $u(z)$ can be considered as solutions of the differential equation of infinite order
\begin{equation}
\left(-\cosh(\delta\, d/dz)-\delta^2/z\right)u(z)=\lambda^{(\delta)} u(z).
\label{eq:model}
\end{equation}
The solutions become uniquely defined if we require that they
coincide with the solutions of the limiting equation
\begin{equation}
-u''(z)/2-u(z)/z=[(\lambda^{(\delta)}-1)/\delta^2] u(z)
\end{equation}
in the limit $\delta\rightarrow 0$. On the positive real
axis this equation is the radial Schr\"odinger equation for the
($l=0$)-states of the hydrogen atom. 

By using the power series ansatz introduced in 
[M. Aunola, J. Math Phys 44, 1913 (2003); math-ph/0207038]
we look for solutions of 
\begin{equation}
-\sum_{m=0}^{\infty}\frac{\delta^{2m}}{(2m)!}\frac{d^{2m}u}{dz^{2m}}-
\frac{\delta^2u}z=\lambda^{(\delta)} u(z)
\end{equation} 
and obtain the leading terms of the solution and eigenvalues.
By reinterpreting the different parts as analytical functions
we find the general form of the solution as

\begin{equation}
u_n^{(\delta)}(z)=\left(\sum_{k=1}^{n}L_k^{(n)}\,\alpha^{(n,\delta)}_k
\, z^k\right)\exp(-z\,\mathrm{arsinh}(\delta/n)/\delta),
\label{eq:gensolu}
\end{equation}
where $\mathrm{arsinh}$ is the inverse of the hyperbolic sine,
$L_k^{(n)}$ are the coefficients of the associated 
Laguerre polynomial of the first kind  and 
the corresponding eigenvalues are given by
\begin{equation}
\lambda_n^{(\delta)}=-\sqrt{1+\delta^2/n^2}.
\end{equation}
The power series expansions converge for $\vert\delta\vert<n$,
but they also apply to all finite, real and positive 
values of $\delta$. 

Simultaneously we diagonalise the infinite matrices $V$ with
elements
\begin{equation}
V_{jj}=-\delta/j,\quad V_{j,j+1}=V_{j+1,j}=-1/2, \quad j=1,2,\ldots
\,.
\label{eq:matrix}
\end{equation}
The $j^{\mathrm{th}}$ component of eigenvector corresponds to
$u_j:=u(\delta j)$. 
In the matrix representation the condition $u_0=0$ has been explicitly
inserted. We can also construct isospectral matrices with
simple eigenvalues, i.e. $\{1/n^{2k}\}_{n=1}^{\infty}$ and,
subsequently, infinite matrices with desired eigenvalues.
In addition, the results of [V. Berezin, Phys. Rev. D 55, 2139 (1997)],
concerning quantum black holes are correctly reproduced. 
The discretised, associated Laguerre polynomials corresponding to
$L_n^1(x)=d L_n(x)/dx$ are defined also.

{\noindent For} \textbf{further details}, see the source file \texttt{Notebook/Exactsolu.nb}
containing all explicit formulae.

{\noindent \textbf{Acknowledgement:} Discussions with Prof. J. Timonen, Prof. O. Civitarese and Dr. L. Kahanp\"a\"a are gratefully acknowledged.
}

\end{document}